\documentclass[11pt]{article}
\usepackage{verbatim,amsmath,eufrak}
\usepackage{cite} % prevents showidx from correct work
\usepackage{hyperref}

\hypersetup{
    bookmarks=true,         % show bookmarks bar?
    unicode=false,          % non-Latin characters in Acrobat�s bookmarks
    pdftoolbar=true,        % show Acrobat�s toolbar?
    pdfmenubar=true,        % show Acrobat�s menu?
    pdffitwindow=false,     % window fit to page when opened
    pdfstartview={FitH},    % fits the width of the page to the window
    pdftitle={My title},    % title
    pdfauthor={Author},     % author
    pdfsubject={Subject},   % subject of the document
    pdfcreator={Creator},   % creator of the document
    pdfproducer={Producer}, % producer of the document
    pdfkeywords={keywords}, % list of keywords
    pdfnewwindow=true,      % links in new window
    colorlinks=false,       % false: boxed links; true: colored links
    linkcolor=red,          % color of internal links
    citecolor=green,        % color of links to bibliography
    filecolor=magenta,      % color of file links
    urlcolor=cyan           % color of external links
}
\renewcommand{\arraystretch}{1.3}
\makeatletter
\newdimen\normalarrayskip              % skip between lines
\newdimen\minarrayskip                 % minimal skip between lines
\normalarrayskip\baselineskip
\minarrayskip\jot
\newif\ifold             \oldtrue            \def\new{\oldfalse}
\def\arraymode{\ifold\relax\else\displaystyle\fi} % mode of array entries
\def\eqnumphantom{\phantom{(\theequation)}}     % right phantom in eqnarray
\def\@arrayskip{\ifold\baselineskip\z@\lineskip\z@
     \else
     \baselineskip\minarrayskip\lineskip2\minarrayskip\fi}
\def\@arrayclassz{\ifcase \@lastchclass \@acolampacol \or
\@ampacol \or \or \or \@addamp \or
   \@acolampacol \or \@firstampfalse \@acol \fi
\edef\@preamble{\@preamble
  \ifcase \@chnum
     \hfil$\relax\arraymode\@sharp$\hfil
     \or $\relax\arraymode\@sharp$\hfil
     \or \hfil$\relax\arraymode\@sharp$\fi}}
\def\@array[#1]#2{\setbox\@arstrutbox=\hbox{\vrule
     height\arraystretch \ht\strutbox
     depth\arraystretch \dp\strutbox
     width\z@}\@mkpream{#2}\edef\@preamble{\halign
\noexpand\@halignto
\bgroup \tabskip\z@ \@arstrut \@preamble \tabskip\z@ \cr}%
\let\@startpbox\@@startpbox \let\@endpbox\@@endpbox
  \if #1t\vtop \else \if#1b\vbox \else \vcenter \fi\fi
  \bgroup \let\par\relax
  \let\@sharp##\let\protect\relax
  \@arrayskip\@preamble}
\def\eqnarray{\stepcounter{equation}%
              \let\@currentlabel=\theequation
              \global\@eqnswtrue
              \global\@eqcnt\z@
              \tabskip\@centering
              \let\\=\@eqncr
 \halign to \displaywidth\bgroup
    \eqnumphantom\@eqnsel\hskip\@centering
    $\displaystyle \tabskip\z@ {##}$%
    \global\@eqcnt\@ne \hskip 2\arraycolsep
         $\displaystyle\arraymode{##}$\hfil
    \global\@eqcnt\tw@ \hskip 2\arraycolsep
         $\displaystyle\tabskip\z@{##}$\hfil
         \tabskip\@centering
    &{##}\tabskip\z@\cr}
\begingroup\ifx\undefined\newsymbol \else\def\input#1 {\endgroup}\fi

\newcounter{app}

\def\app{\setcounter{equation}{0}
\def\theequation{A\Roman{app}.\arabic{equation}}\par
   \addvspace{4ex}
   \@afterindentfalse
  \secdef\@app\@dapp}
\newcommand\@app{\@startsection {app}{1}{0ex}%
                                   {-3.5ex \@plus -1ex \@minus -.2ex}%
                                   {2.3ex \@plus.2ex}%
                                   {\normalfont\Large\bf}}

\def\@dapp#1{%
{\parindent \z@ \raggedright  \bf #1}\par\nobreak}
\def\l@app#1#2{\ifnum \c@tocdepth >\z@
    \addpenalty\@secpenalty
    \addvspace{1.0em \@plus\p@}%
    \setlength\@tempdima{8.5em}%
    \begingroup
      \parindent \z@ \rightskip \@pnumwidth
      \parfillskip -\@pnumwidth
      \leavevmode \bfseries
      \advance\leftskip\@tempdima
      \hskip -\leftskip
      #1\nobreak\hfil \nobreak\hb@xt@\@pnumwidth{\hss #2}\par
    \endgroup\fi}
\newcounter{sapp}[app]

\def\sapp{\def\theequation{A\arabic{app}.\arabic{equation}}\par
   \@afterindentfalse
  \secdef\@sapp\@dsapp}
\newcommand\@sapp{\@startsection{sapp}{2}{\z@}%
                                     {-3.25ex\@plus -1ex \@minus -.2ex}%
                                     {1.5ex \@plus .2ex}%
                                     {\normalfont\large\bfseries}}

\def\@dsapp#1{%
{\parindent \z@ \raggedright  \bf #1}\par\nobreak}
\newcommand{\l@sapp}{\@dottedtocline{2}{1.5em}{3em}}
\def\be{\begin{eqnarray}}
\def\ee{\end{eqnarray}}
\def\nn{\nonumber}

\def\p{\partial}

\def\beq{\begin{equation}}
\def\eeq{\end{equation}}
\def\ba{\beq\new\begin{array}{c}}
\def\ea{\end{array}\eeq}
\def\be{\ba}
\def\ee{\ea}

\def\Tr{{\rm Tr}\,}

\newfont{\alef}{msbm10 at 11pt}
\newfont {\goth}{eufm10 at 11pt}
\def\mathbb#1{\hbox{{\alef #1}}}

\let\@@savethanks\thanks
\def\thanks#1{\gdef\thefootnote{\alph{footnote}}\@@savethanks{#1}}

\bigskip

\title{
\bigskip
{\bf
From Hurwitz numbers to Kontsevich--Witten tau-function: a connection by Virasoro operators} \vspace{.5cm}}
\author{{\bf Alexander Alexandrov}\thanks{E-mail:  {\tt alexandrovsash at gmail.com}}
\date{ } \\
{\small {\it Mathematics Institute, University of Freiburg,
Freiburg, Germany \&}}\\
{\small {\it ITEP, Moscow, Russia}}\\
}

\begin{document}

\setcounter{footnote}{0}

\setcounter{tocdepth}{3}

\maketitle

\vspace{-8.0cm}

\begin{center}
\hfill ITEP/TH-40/11
%\hfill LPT ENS-10/37\\
%\hfill IPHT-t10/146\\
\end{center}

\vspace{6.5cm}
\bigskip

\begin{abstract}
In this letter ,we present our conjecture on the connection between the Kontsevich--Witten and the Hurwitz tau-functions. The conjectural formula connects these two tau-functions by means of the $GL(\infty)$ group element. An important feature of this group element is its simplicity: 
this is a group element of the Virasoro subalgebra of $gl(\infty)$. If proved, this conjecture would allow to derive the Virasoro constraints for the Hurwitz tau-function, which remain unknown in spite of existence of several matrix model representations, as well as to give an integrable operator description of the Kontsevich--Witten tau-function.
%This construction also allows us to represent the Kontsevich--Witten tau-function in terms of $GL(\infty)$ operators.
\end{abstract}

\bigskip


{\small \bf MSC 2010 Primary: 37K10, 14N35, 81R10, 17B68; Secondary: 81T30.}

\begin{comment}
(	37K10  	Completely integrable systems, integrability tests, bi-Hamiltonian structures, hierarchies (KdV, KP, Toda, etc.)
	14N35  	Gromov-Witten invariants, quantum cohomology, Gopakumar-Vafa invariants, Donaldson-Thomas invariants
   53D45  	Gromov-Witten invariants, quantum cohomology, Frobenius manifolds
		81R10  	Infinite-dimensional groups and algebras motivated by physics, including Virasoro, Kac-Moody, $W$-algebras and other current algebras and their representations 
			81R12  	Relations with integrable systems 
				17B68  	Virasoro and related algebras
				22E70  	Applications of Lie groups to physics; explicit representations
				81T30  	String and superstring theories; other extended objects (e.g., branes))
\end{comment}
\bigskip

\bigskip

\bigskip

%\tableofcontents

\def\thefootnote{\arabic{footnote}}

\section*{Introduction}
\def\theequation{\arabic{equation}}
\setcounter{equation}{0}

Generating functions in enumerative geometry constitute a particular subclass of partition functions appearing in string theory. It is commonly assumed that partition functions from this subclass possess nice integrable properties and simple matrix model representations. However, a number of examples for which this type of description is established are relatively small (see e.g. \cite{OPHur,TSII} and references therein). For instance, integrable hierarchies behind the full generating functions of the Gromov--Witten invariants are known only for the simplest compact manifolds (a point and a two dimensional sphere) and some of their deformations and modifications. This is why it is important to investigate the common properties of the known examples and to establish relations between them. In this letter, we conjecture a new, integrable, reformulation of the relation between two important tau-functions belonging to the domain of enumerative geometry: the Kontsevich--Witten tau-function and a generating function of the simple Hurwitz numbers, which we call the Hurwitz tau-function for the simplicity sake.

Since the early 1990s, when Kontsevich's matrix integral \cite{Konts} was used for the description of the intersections on the moduli spaces (two-dimensional topological gravity investigated by Witten \cite{Witten}), the Kontsevich--Witten tau-function became an inevitable part of mathematical physics \cite{GKM}. 
%generating function for the Gromov-Witten invariants of a point
It can arguably be considered as an elementary building block for a huge family of more complicated partition functions.\footnote{However see \cite{OP}, where it is claimed that the partition function of $CP^1$ model should be considered as the most elementary one.}
%Virasoro constraints, which play crucial role in the subject, unambiguously fix Kontsevich--Witten tau-function and follows %from the Kontsevich matrix integral.
As a consequence of its special role, the Kontsevich--Witten tau-function is very well studied. %and a lot of elements of universal description, which are still lacking for more complicated examples, are known in this case. 
In particular, for this tau-function, we know the already mentioned  Kontsevich matrix integral representation, Virasoro constraints, integrable properties, moment variables description, random partitions representation, spectral curve based description and a vast net of relationships with other models (see, e.g., \cite{Morint, KS,KS2,IZ, KontsAMM} and references therein).
%This famous matrix integral
%(in addition to fermionic representation)
%is probably the most convenient expression for the Kontsevich--Witten tau-function.
%The matrix integral representation is very useful in some cases, in particular for understanding of the integrable properties or derivation of the Virasoro constraints. In other cases such representation is not so efficient. The reason is that Kontsevich matrix integral explicitly depends on the external matrix but not on the KdV times, which are Miwa combinations of matrix eigenvalues.

 Counting of the ramified coverings of Riemann surfaces is
another classical subject of enumerative geometry.
 %is the theory of Hurwitz numbers.
%which is deeply connected with the theory of Gromov-Witten invariants of two dimensional sphere \cite{OPHur}.
%Generating functions of them can be expressed as random partition models by classical Frobenius's formula \cite{Dijkgmir}.
Under particular constraints, the number of possible ramifications is finite  and the generating function of these numbers (called the Hurwitz numbers) can be represented in terms of the cut-and-join operator \cite{HHK,HHK1}. This operator belongs to the (central extension of the) $gl(\infty)$ algebra, acting on the space of solutions of the KP hierarchy, what guarantees the KP integrability of the generating function \cite{ammta}. In what follows, we deal only with ramifications over the Riemann sphere. 
%These operators are similar to Givental ones, and they have a clear integrability meaning: they are the symmetries of integrable hierarchy, what means that they map a $\tau$-function to some, in general another, $\tau$-function \cite{Kazarian}.
%While existing examples show that the cut-and-join-like operators are natural in description of standard matrix integrals \cite{Morsh}, establishing of general relations requires an additional investigation.
%This generating function can be restored by topological recursion \cite{MarBouch,HHK5},
%For this generating function at least two different matrix integrals are known: one \cite{EynHur}
%developing the ideas of \cite{EynP}, and another \cite{Morsh} with usual integration contours but non-flat measure, which is a particular case of a more general family of matrix integrals \cite{MMRP}.
Several matrix model representations of the generating functions for both simple (with arbitrary ramification over one point) and double (with arbitrary ramification over two points) Hurwitz numbers are known \cite{MarBouch,EynHur,Morsh,MMRP}.

In this letter, we conjecture a novel relation between the Kontsevich--Witten tau-function and the Hurwitz tau-function. These two functions are very well known to be closely related \cite{HHK2,OPHur,HHK3,MMHodge} and our conjecture is based on these known relations. In particular, the Hurwitz numbers are related to the linear Hodge integrals via the profound Ekedahl--Lando--Shapiro--Vainshtein (ELSV) formula \cite{ELSV}. On the level of generating functions, this relation is given by a $GL(\infty)$ operator, which is a group element of the Virasoro algebra and thus preserves integrability. On the other hand, the generating function of the Hodge integrals is a deformation of the Kontsevich--Witten tau-function
%correspondent Gromov-Witten generating function
%by a Givental operator 
\cite{Mumford,GWHodge,Giv,Giv1}.  Integrable properties of this deformation are more subtle. Namely, the generating function of the Hodge integrals and the Kontsevich--Witten tau-function are connected with each other by a Givental operator, which does not belong to the $GL(\infty)$ group, supplemented by a linear change of variables, which is not equivalent to the local change of the spectral parameter. We conjecture that, while \emph{neither} Givental operator \emph{nor} change of variables respects KP integrability, their combination can be substituted by a simple $GL(\infty)$ operator, which \emph{does} respect the KP integrability. Moreover, this operator is conjectured to be also  a group element of the Virasoro algebra. Hence the $GL(\infty)$ operator connecting two tau-functions is particularly simple: only components of the Virasoro subalgebra of $gl(\infty)$ contribute to it:
\be
Z_K(\tau_k=t_{2k+1})=e^{\widehat{V}_0}\cdot e^{\widehat{V}_+}\cdot e^{\widehat{V}_-}\cdot e^{-\sum_{k>0}\frac{k^{k-1}\beta^{k-1}t_k}{k!g^2}} Z_H(t_k;\beta),
\label{mastcon1}
\ee
where the operators
\be
\widehat {V}_0=a_{0}\log\beta~\widehat{L}_0 \\
\widehat{V}_\pm=\sum_{k>0}a_{\pm k} \beta^{\mp k} \widehat{L}_{\pm k}
\ee
are linear combinations of the Virasoro operators (\ref{virfull}) and $a_{k}$ are rational numbers.
%The change of variables in $Z_K$ corresponds to transition of standard matrix model times to KP times.
Let us stress that the l.h.s. of (\ref{mastcon1}) does not depend both on the parameter $\beta$ and on even times $t_{2k}$.

This class of relations between two KP tau-functions (in our case one of them is a tau-function of the 2-reduced KP hierarchy, that is the KdV hierarchy) was described already in \cite{Orlov}. 
Transformations of this type are the simplest ones among all non-trivial B\"{a}cklund transformations and they naturally generalize relations between equivalent hierarchies \cite{Shiota}. Namely, instead of the group element from the subalgebra consisting of the negative components of the Virasoro algebra, which describes a connection between equivalent hierarchies, here we deal with a group element of the full Virasoro algebra.

%Disadvantage is that the connection with the Hodge intersection numbers, which was of importance in connecting ZK and ZH, disappears from our identities. ???

The organization of this paper is as follows. Sections \ref{K}--\ref{H} are devoted to the brief reminder of the
basic properties of two tau-functions connected by a conjectural relation and the $gl(\infty)$ symmetry algebra of the KP hierarchy. In particular, we focus on the matrix model and the cut-and-join type representations of tau-functions.
%In Section \ref{K} we remind the reader some details about the Kontsevich--Witten tau-function and a construction of the cut-and-join-type representation for it. This representation is not quite satisfactory, because the operator does not belong to $GL(\infty)$ group. In Section \ref{GL} we remind the reader some basic facts about $gl(\infty)$ symmetry algebra of the KP hierarchy.
%In Section \ref{H} we consider the Hurwitz tau-function in terms of the cut-and-join operator and give some related matrix model representations of this function. 
In section \ref{main}, we formulate our conjecture. Section \ref{conc} contains conclusions and open questions.

\section{Kontsevich--Witten tau-function \label{K}}
%In this section we basically follow a matrix model theory point of view and do not give an explanation of the  enumerative geometry origin of the Kontsevich--Witten tau-function, which can be found elsewhere.
%\footnote{Here we slightly change our notations, namely instead of times $\tau_k$ we use KP odd times \be
%t_{2k+1}=\sqrt{2}\tau_k
%\ee
%This set of times was used at \cite{KontsAMM} and makes some formulas a bit simpler.}

%To fix our notations let us remind the reader a definition of the Kontsevich--Witten tau-function. 

Let $\overline{\cal M}_{p;n}$ be the Deligne--Mumford compactification of the moduli space of genus $p$ complex curves $X$ with $n$ marked points $x_1, \dots , x_n$. 
Let us associate with a marked point a line bundle ${\cal L}_i$ whose fiber at a moduli point $(X;x_1,\dots,x_n)$ is the cotangent line to $X$ at $x_i$. 
 Intersection numbers of the first Chern classes of these holomorphic line bundles 
 \be
\int_{\overline{\cal M}_{p;n}}\psi_1^{m_1}\psi_2^{m_2}\dots\psi_n^{m_n}=\langle\sigma_{m_1}\sigma_{m_2}\dots\sigma_{m_n}\rangle
\ee
are rational numbers, not equal to zero only if
\be
\sum^n_{i=1} (m_i-1)=3p-3
\label{dimcon}
\ee
Their generating function
\be\label{Ksum}
F_K(\tau_k)=\left\langle\exp\left(\sum_{m=0}^\infty (2m+1)!! g^{\frac{2(m-1)}{3}}\tau_m\sigma_m\right)\right\rangle\\
=\frac{\tau_0^3}{3!g^2}+\frac{\tau_1}{8}+\frac{\tau_0^3\tau_1}{2g^2}+\frac{5}{8}\tau_0\tau_2+\frac{105}{128}\tau_4g^2+\dots
\ee
is given by the Kontsevich matrix integral over the Hermitian matrix $\Phi$
%\footnote{For review and the list of the references see e.g. \cite{IZ,ammp,Morint}.}
\be
Z_K(\tau_k)=\exp \left(F_K(\tau_k) \right) =\frac{\displaystyle{\int \left[d \Phi\right]\exp\left({\frac{1}{g}\Tr\left(i\frac{\Phi^3}{3!}+\frac{\Lambda \Phi^2}{2}\right)}\right)}}{\displaystyle{\int \left[d \Phi\right]\exp\left({\frac{1}{g}\Tr\frac{\Lambda \Phi^2}{2}}\right)}}
\label{matint}
\ee
This integral depends on the external matrix $\Lambda$, which is assumed to be a negative defined diagonal matrix. This dependence is given by a formal series expansion in time variables $\tau_k$, which are given by the Miwa transform of the matrix $\Lambda$:
\be
\tau_k=\frac{g}{(2k+1)}\Tr{\Lambda^{-2k-1}}
\ee 
%The tau-function depends on the infinite set of the independent time variables $\tau_k$, that means that the size of the matrix tends to infinity.
All $\tau_k$ can be considered as independent variables as the size of the matrices tends to infinity and in this limit (\ref{matint}) gives the Kontsevich--Witten tau-function.

From (\ref{dimcon}) it follows that the parameter $g$ governs the genus expansion:
\be
F_K(\tau_k)=\sum_{p=0}^\infty g^{2p-2}{\cal{F}}^{(p)}_K(\tau_k)
\ee
As we see from (\ref{Ksum}), $g$ is not an independent parameter -- we can get rid of it by the change of variables $\tau_k \mapsto g^\frac{2(1-k)}{3}\tau_k$. 

The Kontsevich--Witten tau-function satisfies an infinite set of linear differential equations known as the Virasoro constraints:
\be
\widehat{\mathcal{L}}_m Z_K=\frac{\p}{\p \tau_{m+1}}Z_K,~~~m\geq-1
\label{vir}
\ee
where the operators
\be
\widehat{\mathcal{L}}_m=\sum_{k=1}^\infty \left(2k+1\right)\tau_k\frac{\p}{\p \tau_{k+m}}+\frac{g^2}{2}\sum_{k=0}^{m-1}\frac{\p^2}{\p \tau_k \p \tau_{m-k-1}}+\frac{\tau_0^2}{2g^2}\delta_{m,-1}+\frac{1}{8}\delta_{m,0},~~~m\geq -1
\label{VirK}
\ee
constitute a subalgebra of the Virasoro algebra:
\be
\left[\widehat{\mathcal{L}}_n,\widehat{\mathcal{L}}_m\right]=2(n-m)\widehat{\mathcal{L}}_{n+m}
\ee
These Virasoro constraints allowed us to derive a cut-and-join type representation for the Kontsevich--Witten tau-function \cite{Koncaj}:
\be
Z_K=e^{\widehat{W}_K} \cdot 1
\label{Mastfor}
\ee
where
\be
\widehat{W}_K=\frac{2}{3}\sum_{k=0}^\infty \left(k+\frac{1}{2}\right)\tau_k\widehat {\mathcal{L}}_{k-1}\\
=\frac{4}{3}\sum_{\substack{k,m\geq 0\\
k+m>0}}\left(k+\frac{1}{2}\right)\left(m+\frac{1}{2}\right)\tau_k\tau_m\frac{\p}{\p \tau_{k+m-1}}\\
+\frac{g^2}{3}\sum_{k,m\geq 0}\left(k+m+\frac{5}{2}\right)
\tau_{k+m+2}\frac{\p^2}{\p\tau_k\p \tau_m}+\frac{1}{g^2}\frac{\tau_0^3}{3!}+\frac{\tau_1}{8}
\label{MastOp}
\ee
The most important disadvantage of this representation is that the operator ${\widehat W}_K$ does not belong to the $gl(\infty)$ algebra, so the integrability is hidden. The main result of this note is a conjecture, which represents the Kontsevich--Witten tau-function in terms of the finite number of $GL(\infty)$ operators acting on the trivial tau-function. This conjectural representation makes integrable properties obvious.

\section{${{ gl}(\infty)}$ algebra \label{GL}}

In this section, we describe a (central extension of the) $gl(\infty)$ algebra, acting on the space of solutions of the KP hierarchy. Actually, we need only $w_{1+\infty}$  subalgebra of $gl(\infty)$. It is generated by the modes of the vertex operator, which due to boson--fermion correspondence are equivalent to the bilinear fermionic operators:\footnote{For more details, see e.g. \cite{JMbook,AZ}}
\be\label{ferm}
\frac{:e^{\widehat{\phi}(w)-\widehat{\phi}(z)}:-1}{w-z}\,\,\,\, \sim \,\,\,\, : \psi(w)\psi^*(z):
\ee
where 
\be
\widehat{\phi}(z)=\sum_{k=1}^\infty \left(\frac{t_k}{g} z^k -g\frac{1}{kz^k} \frac{\p}{\p t_k}\right)
\ee
Normal ordering of the bosonic operator here means that all derivatives stay to the right of all times, for example
\be
:t_i\frac{\p}{\p t_k} t_m:=t_i t_m \frac{\p}{\p t_k}
\ee
The modes of the l.h.s. of (\ref{ferm}) can be represented in terms of the the bosonic current 
\be
\widehat{J}(z)=\frac{\p}{\p z} \widehat{\phi}(z)=\sum_{k=1}^\infty \left(\frac{k}{g} t_k z^{k-1} +\frac{{g}}{z^{k+1}}\frac{\p}{\p t_k}\right)
\label{bascur}
\ee
which itself generates the Heisenberg subalgebra.
%Boson-fermion correspondence: $gl(\infty)$ algebra is generated by the vertex operator 
%\begin{equation}
%\frac{:V_+(w;{\bf t})V_-(z;{\bf t}):-1}{w-z}  \,\,\,\, \sim \,\,\,\, : \psi(w)\psi^*(z):
%\end{equation}
For $w=z+\epsilon$ we have the following expansion
\be\label{decomcur}
\begin{gathered}
\frac{:e^{\widehat{\phi}(z+\epsilon)-\widehat{\phi}(z)}:-1}{\epsilon}=\widehat{J}(z)+\frac{\epsilon}{2}\left(:\widehat{J}(z)^2:+\widehat{J}'(z)\right)\\
+\frac{\epsilon^2}{3!}\left(:\widehat{J}(z)^3:+3:\widehat{J}'(z)\widehat{J}(z):+\widehat{J}''(z)\right)+O(\epsilon^3)
\end{gathered}
\ee
Group elements, corresponding to the first term of this expansion (the current itself) generate the 
time shifts 
\begin{equation}
\exp\left( \sum_{k=1}^\infty a_k \frac{\p}{\p t_k }\right) Z_G({\bf t})=Z_G({\bf t+ a})
\end{equation}
and multiplication of the tau-function by the exponential of a linear function of times 
\begin{equation}
Z_{G}({\bf t})=\exp\left( \sum_{k=1}^\infty b_k  t_k \right) Z_G({\bf t})
\end{equation}
These are obvious symmetries of the tau-functions. 

The next term of the $\epsilon$ expansion, $:\widehat{J}(z)^2:$ (the term $\widehat{J}'(z)$ plays no role), generates the Virasoro algebra with the central charge $c=1$:
\begin{equation}
:\widehat{J}(z)^2:=2 \sum_{m=-\infty}^\infty \frac{\widehat{L}_m}{z^{m+2}} 
\end{equation}
where 
\begin{equation}
\label{virfull}
\widehat{L}_m=\sum_{k=1}^\infty k t_k \frac{\p}{\p t_{k+m}}+\frac{g^2}{2} \sum_{a+b=m} \frac{\p^2}{\p t_a \p t_b} +\frac{1}{2g^2} \sum_{a+b=-m}a\, b\, t_a t_b
\end{equation}
so that
\begin{equation}
[\widehat{L}_k,\widehat{L}_m]=(k-m)\widehat{L}_{k+m}+\frac{1}{12}\delta_{k,-m}(k^3-k)
\end{equation}
Group elements of the Borel subalgebra generated by $\widehat{L}_m$ with $m<0$ are known to connect equivalent KP hierarchies.

%As it follows from the vertex operator representation of the KP hierarchy \cite{JMbook} ???

%$g^2$ plays the role of the genus expansion parameter Takasaki ??? 

%$GL(\infty)$ group is generated by normal ordered powers $:\widehat J(z)^k:$
%of the bosonic current on a sphere
%\be
%\widehat{J}(z)=\sum_{k=1}^\infty \left(k t_k z^{k-1} +\frac{{g^2}}{z^{k+1}}\frac{\p}{\p t_k}\right)
%\label{bascur}
%\ee

%so that the corresponding group elements are given by constant time shift $\exp(\alpha_{-k}\p _{t_{k}})$ and by multiplication by $\exp(\alpha_k t_k)$, which obviously preserve the KP equations. The key role in our construction is played by the Virasoro algebra, which appears at $k=2$:

%\be
%:\widehat{J}(z)^2:=2{g^2}\sum_{k=-\infty}^\infty\frac{\widehat{L}_k}{z^{k+2}}
%\ee
%where
%\be
%\widehat{L}_m=\sum_{k=1}^\infty k t_k \frac{\p}{\p t_{k+m}}+\frac{{g^2}}{2} \sum_{a+b=m} \frac{\p^2}{\p t_a \p t_b} +\frac{1}{2{g^2}} \sum_{a+b=-m}a b t_a t_b
%\label{virfull}
%\ee
%These operators constitute a central extension of the Virasoro algebra with central charge $c=1$ \cite{Fukuma}:
%\be
%\left[\widehat{L}_n, \widehat{L}_m\right]= (n-m)\widehat{L}_{n+m}+ \frac{1}{12}(n^3-n)\delta_{n,-m}
%\ee

The next term of the expansion, up to a certain combination of the previous terms, generates $\widehat{W}_3$ algebra:
\be
:\widehat{J}(z)^3:=3\sum_{k=-\infty}^\infty\frac{\widehat{W}_k}{z^{k+3}}
\ee
an element of which
\be
{\widehat{W}}_0=\sum_{i,j\geq 1} \frac{i j t_it_j}{g}\frac{\p}{\p t_{i+j}} +{g}(i+j)t_{i+j}\frac{\p^2}{\p t_i \p t_j}
\label{CAJ}
\ee
is the cut-and-join operator of the Hurwitz tau-function (see below).

Let us stress that the Virasoro operators (\ref{VirK}) for the Kontsevich--Witten tau-function can be given in terms of the odd part of the current (\ref{bascur}), namely:
\be
\widehat{\mathcal{J}}(z)=\sum_{k=1}^\infty \left(\frac{2k+1}{g} \tau_k z^{2k} +\frac{{g}}{z^{2k+2}}\frac{\p}{\p \tau_k}\right)
\ee
where $\tau_k=t_{2k+1}$, so that
 \be
:\widehat{\mathcal{J}}(z)^2:=2\sum_{k=-\infty}^\infty\frac{\widehat{\mathcal{L}}_k}{z^{2k+2}}-\frac{1}{4z^2}
\ee
%This is the different pre-factor in front of the second term in this current what is partially responsible for the difference between $\tau_{k}$ and $t_{2k+1}$ in (\ref{mastcon}).

\section{Hurwitz tau-function \label{H}}

Hurwitz numbers count ramified coverings of a Riemann sphere. More specifically, the simple Hurwitz number $h(p|m_1,\ldots,m_n)$ gives the number of the Riemann sphere coverings with $N$ sheets, $M$ fixed simple ramification points and a single point with ramification structure given by $\{m_i\}$, a partition of $N$. The number of the double ramifications $M$, the genus $p$ of the cover and the partition $\{m_i\}$ are related:
\be
M=2p-2+\sum_{i=1}(m_i+1)
\ee\label{dimcon2}
One can introduce the generating function of the simple Hurwitz numbers
\be
H(t_k)=\sum_{p=0} g^{2p-2}\sum_{n=1}^\infty \frac{1}{n!} \sum_{m_i;M}\frac{\beta^{M}}{M!}h(p|m_1,\dots,m_n){p_{m_1}\dots p_{m_n}}
\ee
where $p_k=k t_k$.
According to \cite{HHK,HHK1}, this sum can be represented in terms of the cut-and-join operator (\ref{CAJ}):
\be
Z_H(t_k;\beta)=\exp\left(H(t_k) \right)=\exp\left(\frac{g \beta}{2}\widehat{W}_0\right)\cdot\exp\left(\frac{t_1}{{g^2}}\right)\\
%=\sum_{\lambda}d_\lambda \chi_\lambda (t) \exp\left(\frac{\beta C_2}{2}\right) \\
=1+\frac{t_1}{g^2}+\frac{e^{\beta g}}{2} \left(\frac{t_1^2}{2g^4}+\frac{t_2}{g^3}\right)+\frac{e^{-\beta g}}{2} \left(\frac{t_1^2}{2g^4}-\frac{t_2}{g^3}\right)+\dots
\ee

%Here again, as for the Kontsevich--Witten tau-function, we will not give the reader a definition of the Hurwitz tau-function in terms of the enumerative geometry, but we will remind some properties related to the matrix model representations.
%A geometrical definition of the Hurwitz numbers and their generating function can be found elsewhere. Here, to fix our notations, we give a representation of the generating function of the simple Hurwitz numbers in termes of the cut-an-join operator (\ref{CAJ}), namely \cite{HHK,HHK1}:
%\be
%Z_H(t_k;\beta)=\exp\left(\frac{\beta}{2}\widehat{W}_0\right)\cdot\exp\left(\frac{t_1}{{g^2}}\right)
%\ee

For this generating function, several matrix model representations are known, let us mention only the simplest ones. The Hurwitz tau-function can be represented as a Hermitian matrix integral \cite{Morsh}:
\be
Z_H(t_k;\beta)=\int_{N\times N} \left [ d \mu(\Phi) \right] \exp\left(-\frac{1}{2\beta}\Tr \Phi^2 +\Tr \left(e^{\Phi-N\beta/2}\psi\right)\right)
\ee
where the measure of integration $\left [ d \mu(\Phi) \right]$ is non-flat and can be represented in terms of the standard measure on the space of Hermitian matrices as follows:
\be
\left[ d \mu(\Phi) \right]=\sqrt{\det\frac{\sinh\left(\frac{{\Phi}\otimes {1}-{1}\otimes{\Phi}}{2}\right)}{\left(\frac{{\Phi}\otimes {1}-{1}\otimes{\Phi}}{2}\right)}} \left[ d \Phi \right]
\ee
The times $t_k$ are given by the Miwa transform
\be
t_k=\frac{1}{k}\Tr \psi^k
\ee
The same tau-function can be given by an integral over normal matrices \cite{MMRP}:
\be
Z_H(t_k;\beta)= \int_{N\times N} \frac{\left[d Z\right]}{\det\left( {{Z Z^\dagger}}\right)^{N+\frac{1}{2}}}
\exp\left({-\frac{1}{2\beta g}\Tr \log^2 {{Z Z^\dagger}}}+\frac{1}{g}\Tr Z^\dagger+\frac{1}{g}\sum_{k=1}^\infty t_k \Tr { Z}^k\right)
\ee
On can get rid of explicit dependence of $N$ by a simple change of integration variable $Z \to Z^{-1}$. From the identity 
\be
\frac{\left[dZ\right]}{\det(Z Z^\dagger)^{N+1}} =\left[d Z^{-1}\right],
\ee
we have
\be
Z_H(t_k;\beta)= \int_{N\times N} \left[d Z\right]
\exp\left(\frac{1}{g} \Tr \left(-{\frac{1}{2\beta}\log^2 {{Z Z^\dagger}}}-\frac{g}{2} \log {{Z Z^\dagger}}+ (Z^\dagger)^{-1}+\sum_{k=1}^\infty t_k  { Z}^{-k}\right)\right)
\ee
In all presented matrix models, it is assumed that the size of the matrices tends to infinity and unimportant $t_k$--independent factors are omitted.

\section{Relation between two tau-functions \label{main}}

The main result of our ``experimental'' investigation is a conjectural formula connecting two tau-functions (Kontsevich--Witten and Hurwitz) defined in the previous sections. Namely, on the basis of the explicit calculations, we claim that\footnote{We have compared the first 150 terms on both sides of this relation.}
\be
Z_K(\tau_k=t_{2k+1})=\widehat{U}_{KH} \cdot Z_H(t_k;\beta)
\label{mastcon}
\ee
with
\be
\widehat{U}_{KH}=e^{\widehat{V}_0}\cdot e^{\widehat{V}_+}\cdot e^{\widehat{V}_-}\cdot e^{-\sum_{k>0}\frac{k^{k-1}\beta^{k-1}t_k}{k!g^2}}
\label{operU}
\ee
and
\be
\widehat{V}_0=a_{0}\log\beta~\widehat{L}_0\nn\\
\widehat{V}_\pm=\sum_{k>0}a_{\pm k} \beta^{\mp k} \widehat{L}_{\pm k}
\ee
Here, $a_k$ are rational numbers. Let us stress that the genus expansion of two tau-functions connected by this relation is quite different (see (\ref{dimcon}) and (\ref{dimcon2}), respectively), thus the dependence on the variable $g$ is nontrivial. Let us stress that in this formulation the change of 
variables is performed by the Virasoro operators,
thus a problem with non-invertible change of variables does not appear at all.

The first part of the relation between two tau-functions is known and
\be
Z_{Hodge}(t_k)=\exp\left(\sum_{k<0} a_k \beta^{-k} \widehat{L}_{k} \right) \exp\left(-\sum_{k=1}^\infty \frac{k^{k-1}\beta^{k-1}t_k}{k!{g^2}}\right) Z_H(t)\\
=1+\frac{t_{{2}}{\beta}^{3}}{6}+{\frac {{t_{{1}}}^{2}g^2+8\,{t_{{1}}}^{3}+6
\,t_{{3}}{g^2}}{48{g^2}}}{\beta}^{4}+{\frac {1}{120}}\,t_{{2}} \left( 60\,t_{{1}}+{g^2}
 \right) {\beta}^{5}\\
 +{\frac {500\,g^2{t_{{2}}}^{2}+414\,
t_{{3}}{g}^{4}+80\,{t_{{1}}}^{3}g^2+{t_{{1}}}^{2}{g}^{4}+240\,{t_{{1}}}^
{4}+2160\,t_{{1}}t_{{3}}g^2}{1440 g^2}}{\beta}^{6}\\
+{\frac {2\,
t_{{2}}{g}^{6}+21224\,t_{{4}}{g}^{4}+15155\,{t_{{1}}}^{2}t_{{2}}g^2+7000
\,{t_{{1}}}^{3}t_{{2}}+5250\,t_{{2}}t_{{3}}g^2+1400\,t_{{1}}t_{{2}}{g}^{
4}+16800\,t_{{4}}t_{{1}}g^2}{10080 g^2}}{\beta}^{7}\\
+O(\beta^{8})
\ee
is a generating function of the linear Hodge integrals (for more details see an excellent paper by Kazarian \cite{HHK3}). 
 From this relationship, the coefficients $a_k$ for the non-positive $k$ can be extracted. Namely, $a_0=-\frac{4}{3}$, as it follows from the Theorem 2.3 of \cite{HHK3}, while coefficients for positive $k$ can be restored from the equation 
\be
\exp\left( \sum_{k>0} a_{-k} z^{k+1}\frac{\p}{\p z}\right) \cdot z
=\frac{z}{1+z}e^{-\frac{z}{1+ z}}
\ee
The first few coefficients $a_k$ are given in the following table:
\bigskip
\begin{center}
\begin{tabular}{|c|c|c|c|c|c|c|c|c|c|}
\hline
 $k$ & $-1$ & $-2$ & $-3$ & $-4$ & $-5$ & $-6$ & $-7$ & $-8$ & $-9$\\
\hline
  &  &  &  & &  &  &  & &\\
 $\displaystyle{a_k}$ & $\displaystyle{-2}$ & $\displaystyle{-\frac{1}{2}}$ & $\displaystyle{-\frac{1}{2\cdot 3}}$ & $\displaystyle{0}$ & $\displaystyle{\frac{3}{2^3\cdot 5 }}$ & $\displaystyle{\frac{1}{3^2\cdot 5}}$ & $\displaystyle{-\frac{211}{2^3\cdot 3^3 \cdot 7}}$ & $\displaystyle{-\frac{187}{2^3\cdot 3^3 \cdot 7}}$ &$\displaystyle{\frac{21751}{2^7\cdot 3^2\cdot 5\cdot 7}}$\\
   &  &  &  &  &  &  &  & &\\
\hline
\end{tabular}
\end{center}
%\be
%a_{-1}=-2\nn\\
%a_{-2}=-\frac{1}{2}\nn\\
%a_{-3}=-\frac{1}{6}\nn\\
%a_{-4}=0\nn\\
%a_{-5}=\frac{3}{40}\nn\\
%a_{-6}=\frac{1}{45}\nn\\
%a_{-7}=-\frac{211}{1512}\nn\\
%\ee
%\be
%Z_K\left(\tau_k=\frac{t_{2k+1}}{\sqrt{2}}\right)=Q Z_H(t)
%\ee
%where $Q$ is a $GL(\infty)$ operator
The operator $\widehat{V}_+$ is a core element of our conjecture. We have found the coefficients $a_k$ for the positive $k$ by a direct calculation, namely
\bigskip
\begin{center}
\begin{tabular}{|c|c|c|c|c|c|c|c|c|}
\hline
 $k$ & $1$ & $2$ & $3$ & $4$ & $5$ & $6$ & $7$ & $8$ \\
\hline
  &  &  &  & &  &  &  & \\
 $\displaystyle{a_k}$ & $\displaystyle{-\frac{2}{3}}$ & $\displaystyle{\frac{2^2}{3^2\cdot5}}$ & $\displaystyle{-\frac{2}{3^3\cdot5}}$ & $\displaystyle{\frac{11}{3^5\cdot5\cdot7}}$ & $\displaystyle{\frac{11}{3^6\cdot5^2}}$ & $\displaystyle{-\frac{2^4}{3^7\cdot5^2}}$ & $\displaystyle{-\frac{359}{3^8\cdot5^3\cdot7}}$ & $\displaystyle{\frac{13\cdot 3137}{2\cdot3^9\cdot5^3\cdot7\cdot11}}$ \\
   &  &  &  &  &  &  &  & \\
\hline
\end{tabular}
\end{center}
Unfortunately, we do not know any simple recursion relation for the $a_k$ with positive $k$. The only observation we made is that an expression
\be
f(z)=\exp\left(\sum_{k>0} a_k z^{1-k}\frac{\p}{\p z}\right) \frac{z}{1+z}e^{-\frac{z}{1+ z}}
\ee
is even
\be
f(z)=f(-z)
\label{sym}
\ee
This property could be related to the fact that the Kontsevich--Witten tau-function satisfies the KdV equations, that is 2-reduction of the KP hierarchy.
Let us stress that the operator, connecting $Z_K$ and $Z_H$, can be represented in various ways, so most probably the representation (\ref{operU}) can be further simplified.

%\be
%Q=\exp\left(-\log\left(\sqrt{2}\beta^{\frac{4}{3}}\right)L_0\right)\exp\left(\sum_{k>0} a_k \beta^{-k} L_{k} \right)\exp\left(\sum_{k<0} a_{k} \beta^{-k} L_{k} \right) \exp\left(-\sum_{k=1}^\infty \frac{k^{k-1}\beta^{k-1}t_k}{k!{g^2}}\right)
%\ee

%\be
%a_1=-\frac{2}{3}\\
%a_2=\frac{2^2}{3^2\cdot5}\\
%a_3=-\frac{2}{3^3\cdot5}\\
%a_4=\frac{11}{3^5\cdot5\cdot7}\\
%a_5=\frac{11}{3^6\cdot5^2}\\
%a_6=-\frac{2^4}{3^7\cdot5^2}\\
%a_7=-\frac{359}{3^8\cdot5^3\cdot7}
%\ee

The formula (\ref{mastcon}) can be of use for investigation of  both the Kontsevich--Witten tau-function and the Hurwitz tau-function. For the Kontsevich--Witten tau-function it gives an integrable representation in terms of $GL(\infty)$ operators. This relation also could be used for construction of a new matrix model representation in terms of times, contrary to the Kontsevich matrix integral representation, dependent on the external matrix.
For the Hurwitz tau-function, our conjectural relation should allow to derive a set of the Virasoro constraints by a simple conjugation of the constraints for the Kontsevich--Witten tau-function. Namely

\be
\widehat{\mathfrak{L}}_n Z_H(t_k; \beta)=0, ~~~~~~~~~~~~~~n\geq -1
\ee
where
\be
\widehat{\mathfrak{L}}_n={\widehat U}^{-1}_{KH} \left(\widehat{\mathcal{L}}_n -\frac{\p}{\p \tau_{n+1}}\right){\widehat U}_{KH}
\ee

We guess that this transformation can be given in terms of the global current on the Lambert spectral curve.

%then the following Virasoro constraints

%\be
%\tilde{L}_k Z_H(t)=0
%\ee

%where

%\be
%\tilde{L}_k=\exp\left(\frac{u^3}{2}W_0\right) L_k \exp\left(-\frac{u^3}{2}W_0\right)
%\ee

\section{Conclusion and open questions \label{conc}}

In this letter, we present a conjectural relation between two important generating functions of enumerative geometry: the Kontsevich--Witten tau-function and the generating function of the simple Hurwitz numbers. Known and conjectured formulas for these two tau-functions contain three different types of operators significant for modern string theory and enumerative geometry.
%This conjecture, if  proved, should help in unification of three overlapping, but different types of operators, which show themselves in modern string theory/enumerative geometry. 
These three types of operators are the Givental operators, the operators generating symmetries of the integrable hierarchies, and the cut-and-join operators.

Givental operators constitute a special class of operators given by the quantization of quadratic hamiltonians \cite{Giv,Giv1}. These operators correspond to the particular transformations of the bosonic currents on a spectral curve. They allow to express some of non-trivial partition functions of string theory in terms of elementary building blocks, such as Kontsevich--Witten tau-functions, and they are conjectured to be applicable in much more general setup. In particular, the Givental operators naturally appear in the matrix model theory \cite{Hdec,Hdec2,KosOr,IMMM,Hdec1,Hdec3}. However, their relation to integrability
remains poorly investigated (see, however, \cite{Giv3}).

$GL(\infty)$ operators preserve the KP-type integrability of the partition functions, that is why relations in terms of these operators look particularly attractive in the context of integrable systems. They are generated by the powers of the bosonic field current and its derivatives. Some matrix integrals can be related with elementary functions by means of $GL(\infty)$ operators \cite{Morsh,MMRP}.

The cut-and-join operator, appearing in the description of the simple and double Hurwitz numbers, also belongs to the $gl(\infty)$ algebra. However, some other cut-and-join type operators, appearing in the description of the more general Hurwitz partition functions, do not belong to this algebra \cite{HHK4,ammta,IPH2}.

It would be extremely interesting to explain presented conjectural relation in terms of matrix integrals. Since both tau-functions from the conjecture can be represented as matrix integrals dependent on external matrix, it could be possible to derive our relation in terms of the operators, acting on the eigenvalues. Relations of this type should generalize the connections between different solutions of the Generalized Kontsevich Model, given by  equivalent hierarchies (for a review, see \cite{Kharchev}).

%Unfortunately, at the moment we do not know how to prove this conjectural relation, not even how to find all the coefficients $a_k$ in a simple form.

%Givental-type expressions for the double Hurwitz numbers? Higher Hurwitz numbers?

%Is it possible to relate two Grassmanian points, corresponding to the Hurwits and Kontsevich through operators, acting on $z$?

%The operators are simple in terms of eigenvalues, can this fact be used to connect two matrix models?

%It would be of particular interest to represent other decomposition formulas of Givental form, in particular those for matrix models \cite{IMMM,MT} in termes of $GL(\infty)$ operators.

%The relation on the level of matrix integrals looks highly non-trivial. ???

\section*{Acknowledgments}
It is a pleasure to thank Alexei Morozov and Andrei Mironov for useful discussions.
This research was partially supported by the ERC Starting Independent Researcher Grant StG No. 204757-TQFT (K. Wendland PI), by RFBR grants 09-02-93105-CNRSL, 11-01-00962, by Ministry of Education and Science of the Russian Federation
under contract 14.740.11.0081 and Russian Federal Nuclear Energy Agency
under contract H.4e.45.90.11.1059. I am grateful to the organizers of the VII-th International Symposium �Quantum Theory and Symmetries� (QTS�7) for
their hospitality. My participation in this conference was supported by ANR project GranMa "Grandes Matrices Al\'{e}atoires" ANR-08-BLAN-0311-01. I am grateful to the organizers of  the HIM trimester on Integrability in Geometry and Mathematical Physics.

\end{document}